\documentclass[twocolumn]{autart}

\usepackage{graphicx}      
\usepackage{natbib}        

\usepackage[T1]{fontenc}
\usepackage[utf8]{inputenc}
\usepackage{float}
\usepackage{amssymb}
\usepackage{amsmath,color}
\usepackage{subcaption}
\usepackage{stmaryrd}
\usepackage{mathbbol}
\usepackage{enumerate}
\usepackage{mathtools}
\usepackage{paralist}
\usepackage{bm}
\usepackage{tikz}
\usepackage[normalem]{ulem} 
\usepackage{etoolbox}
\usepackage{natbib}

\newcommand*\MyScale{1}

\tikzset{every picture/.style={scale=\MyScale,}}

\usepackage{pgfplots}
\pgfplotsset{compat=newest}
\pgfplotsset{plot coordinates/math parser=false}

\newlength\fheight 
\newlength\fwidth  
\newcommand{\R}{\mathbb{R}}

\newcommand{\N}{\mathbb{N}}
\newcommand{\Z}{\mathcal{Z}}
\newcommand{\U}{\mathcal{U}}
\newcommand{\X}{\mathcal{X}}
\newcommand{\K}{\mathcal{K}}

\newcommand{\zeros}{\mathbb{0}}

\newcommand\numeq[2]%
  {\stackrel{\scriptscriptstyle(\mkern-1.5mu#1\mkern-1.5mu)}{#2}} 
\newcommand{\useq}{\mathbf{u}}
\newcommand{\xseq}{\mathbf{x}}

\DeclareMathOperator*{\argmin}{arg\,min}

\newtoggle{proofIFAC}
\toggletrue{proofIFAC}

\begin{document}
\begin{frontmatter}

\title{Forward-looking persistent excitation in model predictive control}


\author[First]{Sven Br{\"u}ggemann}
\author[First]{Robert R. Bitmead}

\address[First]{Mechanical \&\ Aerospace Engineering Department, University of California, San Diego, CA 92093-0411, USA, (e-mails: \{sbruegge, rbitmead\}@eng.ucsd.edu)}

\begin{abstract}                
This work deals with the problem of simultaneous regulation and model parameter estimation in adaptive model predictive control. We propose an adaptive model predictive control and conditions which guarantee a persistently exciting closed loop sequence by only looking forward in time into the receding prediction horizon. Earlier works needed to look backwards and preserve prior regressor data. Instead, we present a procedure for the offline generation of a persistently exciting reference trajectory perturbing the equilibrium. With the new approach we demonstrate exponential convergence of nonlinear systems under the influence of the adaptive model predictive control combined with a recursive least squares identifier with forgetting factor despite bounded noise. The results are, at this stage, local in state and parameter-estimate space.
\end{abstract}


\end{frontmatter}

\section{Introduction}


This paper revolves around a model predictive control (MPC) framework satisfying conditions for closed loop identification. If the system is such that the control input influences both the system state and its uncertainty (e.g. in the form of the covariance of the parameter or state estimate error), then the control inherits a dual function \citep{Feldbaum}. In this case, on the one hand, the control objective involves the desire for regulation or trajectory tracking and hence a steady or slowly varying state. On the other hand, for identification purposes, to reduce the uncertainty and extract more information from the measurement, the system is to be excited (see e.g. \citet{bar-shalom}). Insufficient excitation may result in bursts of oscillatory behavior through parameter drift (\citet{ANDERSON1985247}), singularity in the information matrix (\citet{mereels}) and an unobservable state even in the scalar case (\citet{brueggemannCDC2019}). Hence, the associated dual feedback control acts as an arbitrator between these antagonistic requirements. An overview of such dual problems can be found in \citet{filatov2000}.

Even though MPC is a widely used technique applied in various industries (e.g. \citet{QIN2003733}), the survey by \citet{MAYNE20142967} points out that the field of adaptive MPC, which relates to the dual problem, has attracted relatively little interest in the controls community. Yet, the idea of using an MPC to fulfill the role of an arbitrator between regulation and excitation has been proposed in different publications. 

One common approach is to impose additional input constraints on the solution of the corresponding optimization problem. In this way, by including past information and thus \emph{looking backwards} in time, the control directly ensures persistence of excitation of the initial step of the MPC solution. For instance, in \citet{genceli1196}, to identify FIR models and drive the related system to a given set point, additional periodic input constraints guarantee a periodic persistently exciting (PE) feedback control. In this way, past and future inputs are evaluated within the optimization. Similarly, \citet{Lu2019} propose a robust tube-based MPC for linear uncertain systems with an additional constraint to provide persistence of excitation. Yet, the closed loop is not guaranteed to be PE. Instead of constraining the entire minimizng control sequence, \citet{marafioti2014} suggest a backward looking memory-based MPC which only constrains the first control input as it is the only element of the sequence which is applied to the system. The control strategy is analyzed for FIR and ARMA models. Feasibility and persistence of excitation can be guaranteed if, among other conditions, the initial control sequence is PE. In a backward looking fashion, \citet{LARSSON20151} take into account the Fisher information matrix generated by past information in a further constraint for the optimization problem and focuses on the implementation of the control scheme.

Instead of modifying the constraints to achieve excitation, a number of authors (\citet{HOVD2004119}, \citet{HEIRUNG201564}, \citet{HEIRUNG2017340}) adapt the cost function so that it also contains the parameter error covariance matrix as a proxy for uncertainty. In this way, the control is \emph{looking forward} to seek persistence of excitation, although, in the light of MPC's receding horizon implementation, a PE property of the closed loop is not immediate. \citet{TANASKOVIC20143019} take a different path and split the dual problem into two. Firstly, a nominal MPC ensures that the constraints hold for any element of a set of possible FIR models. Then, the second stage ensures an exciting property by solving an optimization with the objective to reduce the size of the set. The idea of optimally selecting a model based on measurements is also pursued in \citet{HEIRUNG2019128}, where the cost function incorporates an additional risk of choosing an incorrect model.

In this work, under the assumption of full state feedback and no constraints, rather than looking back using past information as in \citet{genceli1196} and \citet{marafioti2014}, the requirement for a PE input is reformulated as a forward looking condition on the reference trajectory, while still guaranteeing the PE property of the closed loop driven by the MPC.
In this way, the main contribution of this work is that persistence of excitation is guaranteed by solely {looking forward} in time despite the MPC's peculiarity of a receding horizon implementation. 
Further, the optimization problem solved online as part of the MPC framework neither complicates nor alters. Instead, the additional constraint of persistence of excitation is reformulated as an algebraic condition on the reference trajectory. Namely, we present a direct constructive procedure for the offline generation of a periodic PE reference trajectory. This extends \citet{SvenBobIFAC2020} where the existence of such a trajectory was assumed. Moreover, local exponential convergence of the closed loop as well as the parameter estimate is ensured for nonlinear systems of which the full state is available. 

\subsection*{Outline}
These results develop as follows subject to conditions which are specified in place. Section~4 provides a procedure to construct a feasible PE periodic reference trajectory.
\begin{enumerate}[4i)]
\item Find an equilibrium state-and-control tuple $(x_s,u_s)$. 
\item In a neighborhood of this equilibrium, select any length-$M$ input sequence perturbation, $\delta \useq_r^M$. Determine a corresponding state sequence, $\xseq_r^M$, as an explicit function of the initial condition $x_r(0)$. 
\item 
Apply the Implicit Function Theorem to solve $x_r(M)=x_r(0)$ for $x_r(0).$ 
\item Initial state $x_r(0)$ and input $u_s+\delta \useq_r^M,$ applied repeatedly, define a period-$M$ solution of the system.
\item Subject to an output reachability assumption, $\delta \useq_r^M$ may be selected to be persistently exciting.
\end{enumerate}
Section~5 deals with PE reference tracking.
\begin{enumerate}[5i)]
\item Apply reference tracking MPC of \citet{koehler2018ACC} to the period-$M$ reference above. This guarantees exponential convergence to the PE reference from an open set of initial conditions, when there is no state disturbance and known parameter. With a suitably bounded disturbance, the state converges exponentially to a neighborhood of the PE reference.
\item This implies that the closed loop signals are PE with the exact parameter.
\item Exponential stability ensures the conservation of PE with a suitably bounded, time-varying parameter error.
\end{enumerate}
For a system model which is linear in the parameters, Section~6 brings in the parameter estimator; recursive least squares with forgetting factor.
\begin{enumerate}[6i)]
\item With zero disturbance, PE signals yield exponential convergence of the parameter error to zero. 
\item For bounded disturbances, they yield exponential convergence to a neighborhood.
\end{enumerate}
Section~7 draws these two ideas together to obtain a combined estimator and MPC-based controller which maintain persistence of excitation of the closed loop while regulating the state to a neighborhood of its equilibrium.
The simulation examples in the next section confirm the theoretical results and underpin their sufficient nature. 

\section{Problem formulation}
Let the system be
\begin{align}
 x_{k+1}&=f(x_k,u_k)+w_k\label{eq:xbar_k+1}
\end{align}
with $ x_k\in\R^n$ being the state, $u_k\in\R^m$ the input and $w_k\in\R^n$ the disturbance at time $k.$ 
Moreover, the state is fully observed and for some $\bar  w>0$ the disturbance satisfies
\begin{align}\label{ass:w}
|w_k|&\leq \bar w. 
\end{align}
System parameters
$\theta=\begin{bmatrix}
 \theta_1 &  \theta_2 & \dots &  \theta_S
\end{bmatrix}^{\top}\in\R^S$ are assumed unknown. Further, $f$ is linear in $\theta$ and may be written
\begin{align}\label{eq:fAsSum}
f(x_k,u_k,)=f_0(x_k,u_k)+\sum_{j=1}^{S} \theta_jf_j(x_k,u_k),
\end{align}
where $f_j:\R^n\times R^m\to\R^n$ are basis functions so that state recursion \eqref{eq:xbar_k+1} becomes
\begin{align}
x_{k+1}=f_0(x_k,u_k)+\varphi_k^{\top} \theta+w_k,\label{eq:xk+1_varphi}
\end{align}
where the regressor 
\begin{align}\label{eq:regressorMatrix}
\varphi_k^{\top}=\begin{bmatrix}
f_1(x_k,u_k) & f_2(x_k,u_k) & \dots & f_S(x_k,u_k)
\end{bmatrix}\in\R^{n\times S}.
\end{align}
Suppose the following.
\begin{assum}\label{ass:fC2}
$f_j:\R^n\times\R^m\to\R^n, j\in\{0,1,\dots,S\}$, are twice continuously differentiable.\hfill$\triangle$
\end{assum}
In order to present concisely the main problem to be solved, we first require a definition of a PE sequence.
\begin{defn} \label{def:PE}
The sequence $\{x_k, u_k\}$ is said to be persistently exciting (PE) if for some constant $M$ and all $j\in\N_{\geq 0}$ there exist positive constants $\alpha$ and $\beta$ such that
\begin{align*}
0<\alpha I_S \leq \sum_{i=j}^{j+M-1}\varphi_i\varphi_i^{\top}\leq\beta I_S<\infty.
\end{align*}\hfill$\triangle$
\end{defn}
The main problem formulation follows.

\begin{prob} Regulate the state $x_k$ in \eqref{eq:xbar_k+1} while guaranteeing that the closed loop sequence $\{x_k, u_k\}$ is PE.\hfill$\triangle$
\end{prob}
In order to solve this problem, we generate a periodic PE reference trajectory around the steady state to which we wish to regulate the system. Furthermore, we present an MPC to track this periodic reference trajectory rendering the controlled state and corresponding feedback control PE. Simultaneously, a recursive least squares identifier with forgetting factor ensures an accurate parameter estimate of $\theta$.

\section{Preliminaries}\label{seq:preliminaires}
Firstly, we introduce the MPC framework based on the nonlinear MPC in \citet{koehler2018ACC} using the notion of incremental stability. Hereby, we include an assumption on the solution of the corresponding optimization problem which entails a continuous feedback law. The section concludes with an assumption on the system being incrementally stabilizable which leads to exponential convergence of the closed loop to a reachable reference trajectory.

\subsection{The model predictive control framework}
Let the MPC-related reference-tracking cost function be
\begin{align*}
J_N(x_k,\useq^N,k)=\sum_{i=0}^{N-1} l(x_{i|k},u_{i|k},k),
\end{align*}
where $x_{i|k}$ represents the state prediction at time instant $k+i$ given the current state $x_k$. The control input $u_{i|k}$ is denoted accordingly. The control sequence from time $k$ to $N-1$ is written as $\useq^N$, where $N$ represents the finite horizon. The running cost
\begin{align*}
l( x_{i|k},u_{i|k},k)&=| x_{i|k}-x_r(k+i)|_Q^2+|u_{i|k}-u_r(k+i)|_R^2,
\end{align*}
where $Q=Q^{\top}>0, R=R^{\top}> 0$, and $\big(x_r(k+i), u_r(k+i)\big)$ is a given reference state and associated control trajectory at time $k+i$. The MPC framework solves the optimization problem
\begin{align}\label{eq:Vn}
V_N(x_k,k)&=\min_{\useq^N} J_N(x_k,\useq^N,k)\\
s.t. & \quad x_{0|k}=x_k\nonumber\\
& \quad \,x_{i+1|k}=f(x_{i|k},u_{i|k})\nonumber
\end{align}
at every time instant $k$ and applies the first control input $u^{\star}_{0|k}$ of the minimizing sequence $\useq^{\star N}$ to the system in \eqref{eq:xbar_k+1}.
\begin{rem}For clarity in our development, we do not include state or input constraints in our formulation here. They can be added within the local stability framework, c.f. \citet{koehler2018ACC}, but would require tracking their associated assumptions connected with evolution within the interior of the feasible set.\end{rem}
%

\subsection{Continuous feedback law}
In order to render the closed loop robustly convergent to the given reference trajectory, a continuous feedback law as well as a stabilizability condition on the system are needed.
Similarly to \citet{mayne1990}, we assume the Hessian matrix of the cost function to be positive definite.
\begin{assum}\label{ass:posDefHessian}
The minimizing control sequence $\useq^{\star N}$ satisfies
\begin{align}\label{eq:Hessian}
\left.\frac{\partial^2 J_N(x_k,\useq^N,k)}{\partial \useq^N\partial \useq^N}\right\rvert_{x_r(k), \useq^{\star N}, k}>0_{mN\times mN}.
\end{align}\hfill$\triangle$
\end{assum}
Note that for this assumption to hold we require strictly positive definite control weight $R$. 
\begin{lem}\label{lem:contU}
Under Assumptions~\ref{ass:fC2} and \ref{ass:posDefHessian} the feedback control $u^{\star}_{0|k}$ related to the MPC in \eqref{eq:Vn} is continuous in $\theta$ and $x_k$ for a neighborhood of $(x_r(k),\theta)$.\hfill$\triangle$
\end{lem}
\begin{pf}
By Assumption \ref{ass:fC2}, $J_N$ is twice continuous differentiable. Then, with Assumption \ref{ass:posDefHessian}, continuity follows from \citet[Theorem 5.1]{johanson2011}. \hfill$\qed$
\end{pf}
Lemma \ref{lem:contU} ensures that a small change in the parameter or state results only in a small change in the generated control sequence. This relation is essential for the local analysis in later sections. 

\subsection{Local incremental stabilizability}
A reference tracking control law, $u_k=\kappa(x_k,x_r(k),u_r(k))$ for reference state $x_r(k)$ and control $u_r(k)$, is introduced by \citet{koehler2019}. Local incremental stabilizability relates to the existence of such a $\kappa$ and is similar to local exponential stabilizability around the given trajectory.
\begin{assum}\label{ass:localIncrementalStabilizability}\citet[Assumption 1]{koehler2019} There exist a control law $\kappa:\R^n \times \R^n \times \R^m \to \R^m$, a $\delta$-Lyapunov function $V_{\delta}:\R^n \times \R^n \times \R^m\to \R_{\geq0}$ that is continuous in the first argument and satisfies $V_{\delta}(x',x',u')=0 \,\,\,\forall \,\,(x',u')$, and parameters $c_{\delta,l},c_{\delta,u},\delta_{loc},k_{max}>0,\rho\in(0,1)$, such that the following properties hold for all $(x,x',u')$ with $V_{\delta}(x,x',u')\leq \delta_{loc}$:
\begin{align*}
c_{\delta,l}|x-x'|^2\leq V_{\delta}(x,x',u')&\leq c_{\delta,u}|x-x'|^2\\
|\kappa(x,x',u')-u'|&\leq k_{max}|x-x'|\\
V_{\delta}(x^+,x'^+,u'^+)&\leq \rho V_{\delta}(x,x',u'),
\end{align*}
where $x^+=f(x,\kappa(x,x',u'))$ and $x'^+=f(x',\allowbreak u')$.\hfill$\triangle$
\end{assum}
Note that neither the $\delta$-Lyapunov function $V_{\delta}$ nor the control law $\kappa$ is required for the implementation of the MPC but its existence is used for the stability analysis. 

\section{Persistently exciting reference trajectory}\label{sec:PErefTraj}
In this section, for the true $\theta$ value, we show how a persistently exciting reference trajectory around a steady state can be generated. Notice that this removes the very limiting assumption of the existence of such a trajecotgry in \citet{SvenBobIFAC2020}. Towards this goal, we require periodicity and feasibility.
\begin{defn}
A \emph{feasible period-$M$ sequence} $(\xseq_r^M,\allowbreak \useq_r^M)$ for system \eqref{eq:xbar_k+1} with $w_k=0$ satisfies 
\begin{align*}
x_r(k+1)&=f(x_r(k),u_r(k)),\\
x_r((k+1)M)&=x_r(kM),\\
u_r((k+1)M)&=u_r(kM),
\end{align*}
for all $k$.\hfill$\triangle$
\end{defn}
For ease of notation, this period, $M,$ coincides with that in Definition~\ref{def:PE} for persistence of excitation. The next lemma shows that under continuity and reachability assumptions on the system, there exists a feasible period-$M$ reference trajectory in the neighborhood of the steady tuple $(x_s,u_s)$. It further reveals that, by choosing just the control sequence, a corresponding initial state for a period-$M$ sequence is defined in the neighborhood of $x_s$.
\begin{lem}\label{lem:perReachRef}
Given Assumption \ref{ass:fC2}, suppose that there exists a steady state
\begin{align*}
x_{s}=f(x_{s},u_{s}),
\end{align*}
with
\begin{align}\label{eq:AB}
A=\left.\frac{\partial f(x,u)}{\partial x}\right\rvert_{x_{s},u_{s}}, \quad B=\left.\frac{\partial f(x,u)}{\partial u}\right\rvert_{x_{s},u_{s}},
\end{align}
where $(A,B)$ controllable and
\begin{align*}
\lambda_i (A)\neq 1, \,i=1,\dots,n.
\end{align*}
Then, for all $M\geq n$ there exists an open set $\U$ with $\mathbf{u}_s^M\coloneqq \{u_s\}_{k=0}^{M-1}\in\U$ such that for any sequence $\useq_{r}^M\in \U$ there exists a feasible period-$M$ sequence $(\xseq_r^M,\allowbreak \useq_r^M)$ with $x_r(0)=g(\useq_r^M)$, where $g:\R^{Mm}\to\R^n$ is continuously differentiable and $\frac{\partial g(\useq_r^M)}{\partial \useq_r^M}$ is of rank $n$.\hfill$\triangle$
\end{lem}
\begin{pf} Let $(x_s(k),u_s(k))$ the steady state and corresponding control at time $k$. Then, for any positive integer $M$,
\begin{align*}
x_s(M)&=f(x_s(M-1),u_s(M-1))\\
&= f\big(f(x_s(M-2),u_s(M-2)),u_s(M-1)\big)\\
&\,\,\, \vdots\\
&=f\bigg( f\Big(\dots\big( f(x_s(0),u_s(0)),u_s(1)\big),\\
& \qquad \qquad  \dots, u_s(M-2)\Big), u_s(M-1)\bigg)\\
&=x_s(0),
\end{align*}
so that
\begin{align}\label{eq:x_r0}
F(x_s,\mathbf{u}_s^M)&\coloneqq x_s\nonumber  \\
& \quad -f\bigg( f\Big(\dots\big( f(x_s,u_s(0)),u_s(1)\big),\nonumber \\
& \qquad \qquad \,\,\,\,  \dots, u_s(M-2)\Big), u_s(M-1)\bigg)\nonumber \\
&=\zeros.
\end{align}
Partially differentiating $F$ with respect to $x_s(0)$ yields
\begin{align*}
I&-\frac{\partial x_s(M)}{\partial x_s(M-1)} \frac{\partial x_s(M-1)}{\partial x_s(M-2)}  \frac{\partial x_s(M-2)}{\partial x_s(M-3)} \dots \frac{\partial x_s(1)}{\partial x_s(0)}\\
&=I-A^M
\end{align*}
which is invertible by hypothesis and the fact that $\lambda_i(A)^M=\lambda_i(A^{M})$. Thus, with Assumption \ref{ass:fC2}, by the Implicit Function Theorem (\citet{rudin1986principles}), there exists an open set $\U\subset \R^{Mm}$ with $\mathbf{u}_s^M\in\U$ such that the following holds. 
There exists a continuously differentiable function $g:\R^{Mm}\to\R^n$ with $g(\mathbf{u}_s^M)= x_s$ such that for all $\useq_r^M\in\U$
\begin{align}\label{eq:algCondPerRef}
F(g(\useq_r^M),\useq_r^M)=\zeros
\end{align}
and
\begin{align}\label{eq:dg/du}
\frac{\partial g(\mathbf{u}_s^M)}{\partial \mathbf{u}_s^M}=-(I-A^M)^{-1}\frac{\partial F(x_s(0),\mathbf{u}_s^M)}{\partial \mathbf{u}_s^M},
\end{align}
which proves the existence of a period-$M$ reference trajectory as \eqref{eq:algCondPerRef} is the algebraic condition for a feasible period-$M$ reference trajectory.
Now, observe that
\begin{align*}
\frac{\partial F(x_s(0),\mathbf{u}_s^M)}{\partial u_s(k)}&=-\frac{\partial x_s(M)}{\partial x_s(M-1)} \frac{\partial x_s(M-1)}{\partial x_s(M-2)} \dots\\
& \qquad \qquad \quad  \frac{\partial x_s(k+2)}{\partial x_s(k+1)} \frac{\partial x_s(k+1)}{\partial u_s(k)}\\
&=-A^{M-k-1}B,
\end{align*}
so that for the last term in \eqref{eq:dg/du} we have that
\begin{align}\label{eq:dF/du}
\frac{\partial F(x_s(0),\mathbf{u}_s^M)}{\partial \mathbf{u}_s^M}&=\left[\begin{matrix}
\frac{\partial F(x_s(0),\mathbf{u}_s^M)}{u_s(0)} & \dots\end{matrix}\right.\nonumber \\
&\left.  \qquad  \begin{matrix} \frac{\partial F(x_s(0),\mathbf{u}_s^M)}{u_s(1)} & \frac{\partial F(x_s(0),\mathbf{u}_s^M)}{u_s(M-1)}\end{matrix}\right]\nonumber \\
&=-\begin{bmatrix}
A^{M-1}B & A^{M-2}B & \dots & B
\end{bmatrix}.
\end{align}
Thus, if $M\geq n$ and $(A,B)$ controllable the matrix in \eqref{eq:dF/du} is of rank $n$ and so is \eqref{eq:dg/du}.\hfill$\qed$
\end{pf}
\begin{rem}Note that the analysis above applies solely inside $\mathcal U$. Hence, the assumption on the eigenvalues is a statement about $\delta x$ and the property that the only solution of $\delta x=A\delta x$ is $\delta x=0$. That is, apart from $x_s,$ there is no other equilibrium inside $\mathcal U$.\hfill$\triangle$
\end{rem}
\begin{rem}The full rank of the partial derivative of $g$ implies that $x_r(0)$ may be altered to any point in a neighborhood of the equilibrium $x_s$ through a corresponding selection of the related sequence $\delta \useq^M$. In general, this sequence -- and hence the completion of the period-M sequence -- is not unique.\hfill$\triangle$
\end{rem}
Note that provided any period-$M$ control sequence $\useq_r^M\in\U$, equation \eqref{eq:x_r0} can be solved numerically in order to compute the initial reference state $x_r(0)$. Given the existence result of a feasible period-$M$ reference trajectory, it is our interest to establish conditions which guarantee its persistence of excitation. Therefore, we continue with the analysis of the system in \eqref{eq:xbar_k+1} linearized at the steady tuple $(x_s,u_s)$ for several reasons: firstly, we want to regulate the system to the steady tuple around which the linearized dynamics are sufficiently accurate; secondly, the result in Lemma \ref{lem:perReachRef} is already local due to the use of the Implicit Function Theorem; thirdly, the analysis of a linear system is simpler and makes the results more intuitive.

Without loss of generality, let $f_0(x_k,u_k)=0$, and denote the corresponding linearized dynamics with $w_k=0$ as
\begin{align}\label{eq:linSys}
\delta x_{k+1}&=A\delta  x_k+B \delta  u_k\nonumber \\
&=\sum_{j=1}^{S} \theta_j A_j\delta   x_k+\sum_{j=1}^{S} \theta_j B_j \delta  u_k,
\end{align}
where
\begin{align}\label{eq:AiBi}
A_j=\left.\frac{\partial f_j(x,u)}{\partial x}\right\rvert_{x_{s},u_{s}}, \quad B_j=\left.\frac{\partial f_j(x,u)}{\partial u}\right\rvert_{x_{s},u_{s}}.
\end{align}
Then, for \eqref{eq:regressorMatrix}, the linearized regressor
\begin{align}\label{eq:linRegressor}
\delta \varphi_k&=\left[\begin{matrix}\delta y_{1}(k) & \delta y_2(k) & \dots & \delta y_{n}(k)\end{matrix}\right],
\end{align}
where $\delta y_{i}(k)=C_{i}\delta  x_k+D_{i}\delta  u_k\in\R^S,\,i\in\{1,2,\dots,n\}$, with 
\begin{align}\label{eq:CsDs}
C_{i}=\begin{bmatrix}(A_1)_{i} \\ (A_2)_{i} \\ \vdots \\ (A_S)_{i}\end{bmatrix}, \quad D_{i}=\begin{bmatrix}(B_1)_{i} \\ (B_2)_{i} \\ \vdots \\ (B_S)_{i}\end{bmatrix}.
\end{align}
Here, $(A_j)_{i}$ represents the $i$th row of matrix $A_j$. Note that for the linearizations above we are explicitly using Assumption \ref{ass:fC2}. 
Our goal is to reformulate the condition of persistence of excitation on the regressor matrix as a condition on the input sequence, whose design we can freely choose. Toward this end, we interpret each column $\delta y_{i}(k)$ of the regressor $\delta \varphi_k$ in \eqref{eq:linRegressor} as one output vector. We analyze each of them with respect to the notion of output reachability defined as follows.
\begin{defn}\label{def:outputReach}(\citet{GREEN1986351})
The system $(A,B,C_i,D_i),\,i\in\{1,2,\dots,n\},$ in \eqref{eq:linSys} and \eqref{eq:CsDs} is said to be \emph{output reachable} if, for any $y\in\R^S$ and arbitrary initial state, there exists an input sequence $\bar u_l,\,l\in\{0,1,\dots,k<\infty\}$ such that the output at time $k$, $\bar y_{i}(k)$, satisfies $\bar y_{i}(k)=y$.\hfill$\triangle$
\end{defn}
Note that in \eqref{eq:CsDs} each $C_i$ and $D_i$ only incorporates row $i$ of matrices $A_j, B_j, \,j\in\{1,\dots,S\}$. Accordingly, each $\delta y_i(k)$ is directly related to only element $i$ of the state vector $\delta  x_k$, denoted by $\delta  x_k^{(i)}$, i.e. with \eqref{eq:linSys},
\begin{align}
\delta  x_{k+1}^{(i)}&=\sum_{j=1}^{S} \theta_j ((A_j)_i\delta   x_k+(B_j)_i \delta  u_k)\nonumber \\
&=\theta^{\top}\delta y_i(k).\label{eq:linSys(i)}
\end{align}
Thus, output reachability may also be viewed as a way to determine if $\delta  x_{k+1}^{(i)}$ is affected by the whole parameter vector $\theta$. 
\begin{rem}
Definition \ref{def:outputReach} can be easily verified. Denoting the McMillan degree of system $(A,B,C_i,D_i)$ as $d_i,\,i\in\{1,2,\dots,n\}$, it is shown in \citet{wolovich} that $(A,B,C_i,D_i)$ is output reachable if and only if
\begin{align*}
\begin{bmatrix}
D_i & C_iB & C_iAB&\dots & C_iA^{d_i-1}B
\end{bmatrix}
\end{align*}
has full rank $S$.\hfill $\triangle$
\end{rem}
Now, we can use the results from \citet{GREEN1986351} to guarantee a PE sequence $\{\delta x_k,\allowbreak \delta u_k\}$ given a persistence of excitation condition on the input sequence only.
\begin{prop}\label{prop:uPEimpliesPhiPE}
Consider the linear system in \eqref{eq:linSys} and let Assumption \ref{ass:fC2} hold. Let $d_i$ the McMillan degree of $(A,B,C_i,D_i),\,i\in\{1,2,\dots,n\}$. Then, the period-$M$ sequence $\{\delta x_r(k),\allowbreak \delta u_r(k)\}$ is PE if there exists an $i$ such that $(A,B,C_i,D_i)$ is output reachable, and if for all $j$ there exist positive constants $\alpha_u$ and $\beta_u$ such that
\begin{align}\label{eq:PE_u}
0<\alpha_u I \leq \sum_{k=j}^{j+M-1-d_i}\delta u_r(k)\delta u_r(k)^{\top}\leq\beta_u I<\infty.
\end{align}\hfill$\triangle$
\end{prop}
\begin{pf}
We consider the linear dynamics in \eqref{eq:linSys} and \eqref{eq:CsDs}. Note that for a period-$M$ sequence $\{\delta x_r(k),\delta u_r(k)\}$ to be PE,
\begin{align}\label{eq:prfSumPhiPhi'}
 \sum_{k=j}^{j+M-1}\delta \varphi_k\delta \varphi_k^{\top}&= \sum_{k=j}^{j+M-1}\delta y_1(k)\delta y_1(k)^{\top} \nonumber \\
 &\qquad + \sum_{k=j}^{j+M-1}\delta y_2(k)\delta y_2(k)^{\top} \nonumber \\
&\qquad  + \dots+\sum_{k=j}^{j+M-1}\delta y_n(k)\delta y_n(k)^{\top}
\end{align}
must be upper bounded and positive definite for all $j$. The upper bound holds trivially by periodicity of the sequence and $f_l,l\in\{1,\dots,S\}$, being twice continuously differentiable (Assumption \ref{ass:fC2}). For the lower bound, if there exists at least one $i\in\{1,\dots,n\}$ for which $\sum_{k=j}^{j+M-1}\delta y_i(k)\delta y_i(k)^{\top}$ is positive definite, \eqref{eq:prfSumPhiPhi'} is positive definite, too, and hence, the corresponding sequence $\{\delta x_r(k),\delta u_r(k)\}$ is PE. By the hypothesis of output reachability of at least one $(A,B,C_i,D_i)$ and the excitation property of the input sequence in \eqref{eq:PE_u}, positive definiteness of at least one summand in \eqref{eq:prfSumPhiPhi'} directly holds by \citet[Corollary 2.1]{GREEN1986351}.\hfill$\qed$
\end{pf}
\begin{rem}
It may be unexpected that we obtain persistence of excitation given a constraint only on one output vector $\delta y_i(k)$. However, it becomes more intuitive if we look at \eqref{eq:linSys(i)}, where we observe that each output vector directly relates to the full parameter vector $\theta$.
\hfill $\triangle$
\end{rem}
As we retain full discretion in selecting the input sequence, the results in this section show how to generate a PE period-$M$ reference trajectory: first select the steady tuple to which the closed loop is desired to converge; after, choose the reference control sequence with PE properties; then numerically solve the algebraic equation in \eqref{eq:x_r0} in order to find the corresponding initial reference state $x_r(0)$. The presentation continues with the controller framework.

\section{Persistently exciting reference tracking with noise} 
The results on persistence of excitation of the closed loop in this section rely on practical stability of the tracking error. As a short intermezzo, we thus first adapt \citet{koehler2018ACC} related to state regulation to the case of trajectory tracking. Then, we analyze persistence of excitation of the closed loop sequence.
It is shown that under the assumption of precise knowledge of the true parameter and a sufficiently small disturbance, the closed loop sequence is PE for all initial conditions $x_0$ within a neighborhood of the initial reference trajectory. Then, an equivalent guarantee is obtained when, additionally, the time-varying uncertain parameter lies within a neighborhood of the actual parameter.

\subsection{Practically stable tracking error}
The following lemma shows exponential convergence of the closed loop to a neighborhood of the reference trajectory, where the size of the neighborhood depends on the bound on the disturbance $w_k$.
\begin{lem}\label{lem:practicalStabilityTrueParameter}
Suppose that Assumption \ref{ass:localIncrementalStabilizability} is satisfied. For any $c_x>0$ there exist $\bar w>0$ and a sufficiently large horizon $N$, such that for all initial conditions $|x_0-x_r(0)|\leq c_x$ and all disturbances $|w_k|\leq \bar w$, the perturbed closed loop converges exponentially to the set $\Z_{RPI}\coloneqq\{x_k-x_r(k):V_N(x_k,k)\leq V_{RPI}(\bar w; N, c_x)\}$, where $V_{RPI}$ is a $\K$-function in $\bar w$ which depends on $N$ and $c_x$.\hfill$\triangle$
\end{lem}
\iftoggle{proofIFAC}{
\begin{pf}
This lemma is concise version of \citet[Theorem 8]{koehler2018ACC} related to state regulation applied to the case of reference tracking. The proof is analogous considering the $\delta$-Lyapunov function from Assumption \ref{ass:localIncrementalStabilizability}.\hfill $\qed$
\end{pf}
}
{
\begin{pf}
This lemma is concise version of \citet[Theorem 8]{koehler2018ACC}. Hence, we adapt the proof of \citet[Theorem 8]{koehler2018ACC} related to state regulation to the case of reference tracking. Therein, Part 1 is satisfied by replacing $|x_k|_Q^2$ by $|x_k-x_r(k)|_Q^2$ and using the upper bound for $V_N(x_k,\theta,k)\leq \gamma |x_0-x_r(0)|_Q^2$ from \citet[Proposition 2]{koehler2019}. Part 2 is invoked as (i) \citet[Proposition 5]{koehler2018ACC} and (ii) \citet[Assumption 4]{koehler2018ACC} hold for the following reasons.

For (i), the proof of the proposition holds by replacing the $\delta$-Lyapunov function therein with that of Assumption \ref{ass:localIncrementalStabilizability} above.

For (ii), the $\delta$-Lyapunov function from Assumption \ref{ass:localIncrementalStabilizability} can directly be used. Therefore, let $\X_f(k)=\{x_f(k):V_{\delta}(x_f(k),x_r(k),u_r(k))\leq \delta_{loc}\}$ and apply the control law $\kappa$ from Assumption \ref{ass:localIncrementalStabilizability}. Then, if $x_f(k)\in\X_f(k)$, by Assumption \ref{ass:localIncrementalStabilizability}, 
\begin{align*}
|x_f(k)-x_r(k)|^2&\leq \frac{1}{c_{\delta,l}} V_{\delta}(x_f(k),x_r(k),u_r(k))\\
|\kappa(x_f(i),x_r(i),u_r(i))-u_r(i)|^2&\leq k^2_{max}|x_f(k)-x_r(k)|^2,
\end{align*}
so that recursively
\begin{align*}
\sum_{i=k}^{\infty}|x_f(i)-x_r(i)|^2&\leq \frac{1}{c_{\delta,l}}V_{\delta}(x_f(k),x_r(k),u_r(k))\sum_{i=k}^{\infty}\rho^k\\
&= \frac{1}{c_{\delta,l}(1-\rho)}V_{\delta}(x_f(k),x_r(k),u_r(k))
\end{align*}
where $x_f(k+1)=f(x_f(k),\kappa(x_f(k),x_r(k),u_r(k)))$, and 
\begin{align*}
\sum_{i=k}^{\infty}|\kappa(x_f(i),&x_r(i),u_r(i))-u_r(i)|^2\\
&\leq k^2_{max}\sum_{i=k}^{\infty} |x_f(i)-x_r(i)|^2 \\
&\leq \frac{k^2_{max}}{c_{\delta,l}(1-\rho)}V_{\delta}(x_f(k),x_r(k),u_r(k)).
\end{align*}
Thus,
\begin{align*}
\sum_{i=k}^{\infty}&l(x_f(i),\kappa(x_f(i),x_r(i),u_r(i)))\\
&\leq \frac{(\lambda_{max}(Q)+\lambda_{max}(R))(1+k_{max}^2)}{c_{\delta,l}(1-\rho)}\\
&\quad \quad \quad V_{\delta}(x_f(k),x_r(k),u_r(k))\\[1em]
&\coloneqq V_f(x_f(k),k)\leq \gamma|x_f(k)-x_r(k)|^2_Q,
\end{align*}
where $\gamma= \frac{c_{\delta,u}(\lambda_{max}(Q)+\lambda_{max}(R))(1+k_{max}^2)}{c_{\delta,l}(1-\rho)}$. The last inequality is satisfied by Assumption \ref{ass:localIncrementalStabilizability}. Lastly, let $c\coloneqq\frac{\delta_{loc}\lambda_{min}(Q)}{c_{\delta,u}}$, where $\lambda_{min}(\cdot)$ denotes the smallest eigenvalue. Then, $|x_f(k)-x_r(k)|_Q^2<c$ implies that $V_{\delta}(x_f(k),x_r(k),\allowbreak u_r(k))\leq \delta_{loc}$, i.e. $x_f(k)\in\X_f(k)$. Consequently, \citet[Assumption 4]{koehler2018ACC} is satisfied. Part 3 of the proof holds directly.\hfill $\qed$
\end{pf}
}
\begin{rem}
A similar convergence result including tightened constraints on the state and control input can be found in \citet{koehler2019b}. In our a case, constraints would make the upcoming statements dependent on an additional condition on the solution being in the interior of the tightened constraint sets.\hfill$\triangle$
\end{rem}

\subsection{Persistently exciting perturbed solution}
Lemma \ref{lem:practicalStabilityTrueParameter} above establishes practical stability of the tracking error in the presence of a bounded disturbance. If the neighborhood of the feasible PE reference trajectory, to which the closed loop converges, is sufficiently small, the corresponding closed loop is PE in finite time.
\begin{lem}\label{lem:PEpracticalStabilityAfterK}
Suppose Assumption \ref{ass:fC2}, \ref{ass:posDefHessian} and \ref{ass:localIncrementalStabilizability} hold and the parameter $\theta$ is known. Then, for a feasible PE reference trajectory and for any $c_x>0$ there exist $\bar w>0$ and a sufficiently large horizon $N$, such that for all $|x_0-x_r(0)|\leq c_x$, $|w_k|\leq \bar w$ and horizon $N$, the closed loop sequence is PE for all $k\geq k_{PE}$ for some $k\in\N_{\geq 0}$.  If $x_0=x_{\bar k}$, where $\bar k\in\{k\in\N_{\geq 0}:k\geq k_{PE}, k\mod M=0\}$, then this holds for all $k$.\hfill $\triangle$
\end{lem}
\begin{pf}
\iftoggle{proofIFAC}
{The proof is based on continuity arguments and divided into three steps. Step I shows the existence of a PE sequence in the neighborhood of the PE reference trajectory. Step II relates $\Z_{RPI}$ to this neighborhood and step III proves the statement using Lemma \ref{lem:practicalStabilityTrueParameter}.

\emph{Step I}: By Assumption \ref{ass:fC2} and \ref{ass:posDefHessian} (via Lemma \ref{lem:contU}), $f$ and $u^{\star}_{0|k}$ are continuous. Thus, for a PE sequence $\{x_r(k),u_r(k)\}$ there exists a positive $\epsilon_{PE}$ such that $|x_r(k)-x_k|\leq \epsilon_{PE}$ implies $\{x_k,u^{\star}_{0|k}\}$ is PE. 

\emph{Step II}: Note that $V_N(x_k,k)<\epsilon$ implies $|x_k-x_r(k)|_Q^2<\epsilon$ for all $\epsilon>0$. Hence, by continuity of $V_{RPI}$, for any $\epsilon_{PE}$ from step I there exist $\epsilon_1, \bar w_1>0$ such that $\min_{\bar x\in\Z_{RPI}}|x-\bar x|_Q^2<\epsilon_1$ implies that $|x-x_r(k)|<\epsilon_{PE}$.

\emph{Step III}:
For any $c_x>0$, let $N$ and $\bar w_2$ satisfy Lemma \ref{lem:practicalStabilityTrueParameter} (requiring Assumption \ref{ass:localIncrementalStabilizability}), and define $\bar w =\min\{\bar w_1,\bar w_2\}$. The statement to be proven holds for $c_x, N, \bar w$ and $k_{PE}$, where $k_{PE}$ is such that $\min_{\bar x\in\Z_{RPI}}|x_{k_{PE}}-\bar x|^2<\epsilon_1$, by using step I and II. The related guarantee for all $k$ is a direct consequence.
}
{}
\hfill$\qed$
\end{pf}

Lemma \ref{lem:PEpracticalStabilityAfterK} establishes that the closed loop sequence is PE given a known parameter $\theta$. As this feature is only of interest if the parameter $\theta$ is unknown and thus, to be estimated, we aim to establish similar results for the case of a time-varying estimate $\hat \theta_k$ within the neighborhood of the true parameter.

\subsection{Persistently exciting perturbed uncertain solution}
\begin{cor}\label{thm:PEAfterKEpsilonTheta}
Suppose Assumption \ref{ass:fC2}, \ref{ass:posDefHessian} and \ref{ass:localIncrementalStabilizability} hold. Let the control input be derived by the optimization problem in \eqref{eq:Vn} with $\theta$ substituted by some $\hat \theta_k, \,k\in\N_{\geq 0}$. Then, for any feasible PE reference trajectory there exist $c_x, \bar w, c_{\theta}>0$ and a sufficiently large horizon $N$ such that for all $|x_0-x_r(0)|\leq c_x$, $|w_k|\leq \bar w$ and $|\theta-\hat \theta_k|\leq c_{\theta}$, the closed loop sequence is PE for all $k\geq k_{PE}$. If $x_0=x_{\bar k}$, where $\bar k\in\{k\geq k_{PE} \wedge k\mod M=0\}$, then this holds for all $k$.\hfill $\triangle$
\end{cor}
\begin{pf}
Let $\hat u^{\star}_{0|k}$ be the MPC feedback control for some $\hat \theta_k$, and $u^{\star}_{0|k}$ that corresponding to $\theta$. Then
\begin{align*}
x_{k+1}&=f(x_k,\hat u^{\star}_{0|k})+w_k\\
&=f(x_k,u^{\star}_{0|k})+\hat w_k,
\end{align*}where
\begin{align*}
\hat w_k=\left(f(x_k,\hat u^{\star}_{0|k})-f(x_k,u^{\star}_{0|k})\right)+w_k.
\end{align*}
By continuity of the MPC feedback via Lemma \ref{lem:contU} and continuity of $f$ through Assumption \ref{ass:fC2}, the disturbance $\hat w_k$ can be bounded by having $\hat \theta_k$ sufficiently close to $\theta$ for all $k$. Thus, the result follows from Lemma \ref{lem:PEpracticalStabilityAfterK}.
\hfill $\qed$
\end{pf}

Corollary \ref{thm:PEAfterKEpsilonTheta} demonstrates that given sufficient assumptions, the closed loop under the influence of the MPC delivers a PE closed loop sequence. It is now time to elaborate why persistence of excitation is desired and, therefore, introduce the estimation algorithm.
\section{Recursive least squares with forgetting factor}
In order to estimate the unknown parameter $\theta$, we select a recursive least squares algorithm with forgetting factor. Therefore, define
\begin{align*}
\tilde x_{k+1}=x_{k+1}-f_0(x_k,u_k),
\end{align*}
and consider the corresponding recursive algorithm
\begin{align}\label{eq:thetaHatRecursion}
\hat \theta_{k+1}=\hat \theta_{k}+P_{k-1}\varphi_{k}D_k^{-1}\left(\tilde x_{k+1}-\varphi_{k}^{\top}\hat \theta_{k}\right),
\end{align}
where $D_k=\lambda T+\varphi_{k}^{\top}P_{k-1}\varphi_{k}$ with $T=T^{\top}>0\in\R^{n\times n}$, and
\begin{align}\label{eq:Pk_recursion}
P_{k+1}&=\lambda^{-1}\Big(I-P_{k}\varphi_{k+1}D_{k+1}^{-1}\varphi_{k+1}^{\top}\Big)P_{k}.
\end{align}
where the forgetting factor $\lambda\in(0,1)$ is constant and $P_{-1}\in\R^{S\times S}$ is symmetric positive definite. The matrix $T$ is related to the weight associated with the prediction error of each element of the state, see the following lemma.
\begin{lem}\label{lem:RLScostFun}
The algorithm in \eqref{eq:thetaHatRecursion} and \eqref{eq:Pk_recursion} converges to the value $\theta$ which minimizes
\begin{align*}
\lambda^{k}|\hat \theta_0- \theta|_{P_{-1}^{-1}}^2+\sum_{i=1}^k \lambda^{k-i}|\tilde x_i-\varphi^{\top}_{i-1} \theta|_{T^{-1}}.
\end{align*}.\hfill$\triangle$
\end{lem}
\begin{pf}
\iftoggle{proofIFAC}
{
The proof is analogous to that of \citet[Theorem 2]{bernstein2019} and hence omitted for brevity.
}
{
Here, we consider the multiple output case. Let
\begin{align*}
L(\theta)&=\frac{1}{2} \sum_{i=1}^k\lambda^{k-i}|\tilde x_i-\varphi_{i-1}^{\top}\theta|^{2}_{T^{-1}}\\
&=\frac{1}{2} \sum_{i=1}^k \lambda^{k-i}\left(\tilde x_i^{\top}T^{-1}\tilde x_i-2\theta^{\top}\varphi_{i-1}T^{-1}\tilde x_i\right.\\
& \quad \quad \quad \quad \quad \quad \left. +\theta^{\top}\varphi_{i-1}T^{-1}\varphi_{i-1}^{\top}\theta\right).
\end{align*}
Differentiating this yields
\begin{align*}
\frac{\partial L(\theta)}{\partial \theta}&=\sum_{i=1}^k \lambda^{k-i}\left(\varphi_{i-1}T^{-1}\varphi_{i-1}^{\top}\theta-\varphi_{i-1}T^{-1}\tilde x_i\right)\\
&=\sum_{i=1}^k \lambda^{k-i}\varphi_{i-1}T^{-1}\left(\varphi_{i-1}^{\top}\theta-\tilde x_i\right)\\
&=\left(\sum_{i=1}^k \lambda^{k-i}\varphi_{i-1}T^{-1}\varphi_{i-1}^{\top}\right)\theta-\sum_{i=1}^k \lambda^{k-i}\varphi_{i-1}T^{-1}\tilde x_i.
\end{align*}
Denote $\hat \theta_k$ as the value of $\theta$ which satisfies the equation above set to zero. Then, one obtains
\begin{align}\label{eq:theta minimizer}
\hat \theta_k=\underbrace{\left(\sum_{i=1}^k \lambda^{k-i}\varphi_{i-1}T^{-1}\varphi_{i-1}^{\top}\right)^{-1}}_{\coloneqq P_{k-1}}\sum_{i=1}^k \lambda^{k-i}\varphi_{i-1}T^{-1}\tilde x_i,
\end{align}
where the matrix $P_{k-1}^{-1}$ can be written in a recursive manner:
\begin{align}\label{eq:PinvRecursion}
P_{k-1}^{-1}&=\sum_{i=1}^k \lambda^{k-i}\varphi_{i-1}T^{-1}\varphi_{i-1}^{\top}\nonumber\\
&=\lambda P^{-1}_{k-2}+\varphi_{k-1}T^{-1}\varphi_{k-1}^{\top}.
\end{align}
Using the Woodbury matrix identity leads to the following recursive expression:
\begin{align}\label{eq:Precursion}
P_{k-1}&=\lambda^{-1}P_{k-2}\nonumber\\
&\quad -\lambda^{-1}P_{k-2}\varphi_{k-1}\left(T+\varphi_{k-1}^{\top}\lambda^{-1}P_{k-2}\varphi_{k-1}\right)^{-1}\nonumber\\
&\quad \varphi^{\top}_{k-1}\lambda^{-1}P_{k-2}\nonumber\\
&=\lambda^{-1}\left(I \vphantom{\left(T^{\top} \right)^{-1}} \right.\nonumber\\
&\quad\left.-\lambda^{-1} P_{k-2}\varphi_{k-1}\left(T+\varphi_{k-1}^{\top}\lambda^{-1}P_{k-2}\varphi_{k-1}\right)^{-1}\varphi_{k-1}^{\top}\right)\nonumber\\
&\quad P_{k-2}\nonumber\\
&=\lambda^{-1}\Big(I \nonumber\\
&\quad-P_{k-2}\varphi_{k-1}\underbrace{\left(\lambda T+\varphi_{k-1}^{\top}P_{k-2}\varphi_{k-1}\right)^{-1}}_{\coloneqq D_{k-1}^{-1}}\varphi_{k-1}^{\top}\Big)P_{k-2},
\end{align}
which is equivalent to \eqref{eq:Pk_recursion}.

Toward the goal of a recursive expression for the parameter vector $\hat \theta_k$, consider
\begin{align*}
\hat \theta_k&\numeq{\ref{eq:theta minimizer}}{=}P_{k-1}\sum_{i=1}^k \lambda^{k-i}\varphi_{i-1}T^{-1}\tilde x_i\nonumber\\
&=P_{k-1}\left(\sum_{i=1}^{k-1} \lambda^{k-i}\varphi_{i-1}T^{-1}\tilde x_i + \varphi_{k-1}T^{-1}\tilde x_k\right)\nonumber\\
&\numeq{\ref{eq:theta minimizer}}{=} P_{k-1}\left(\lambda P_{k-2}^{-1}\hat \theta_{k-1}+ \varphi_{k-1}T^{-1}\tilde x_k\right)\nonumber\\
&\numeq{\ref{eq:Precursion}}{=}\hat \theta_{k-1}+P_{k-2}\varphi_{k-1}D_{k-1}^{-1}\left(\tilde x_k-\varphi_{k-1}^{\top}\hat \theta_{k-1}\right),
\end{align*}
which derives \eqref{eq:thetaHatRecursion}.
}
\hfill$\qed$
\end{pf}

We wish to obtain convergence of the estimate to (a neighborhood of) the true parameter, or equivalently a converging estimation error
\begin{align}\label{eq:thetaTilde}
\tilde \theta_{k}&=\theta-\hat \theta_{k}.\\
\end{align}
This is achieved by the next lemma, whose sufficient condition underpins our desire for a PE closed loop. The result is an extension of \citet{JOHNSTONE1982} adapted to multiple output systems.

\begin{lem}\label{lem:tilde theta convergence}
Suppose the sequence $\{x_k, u_k\}$ is PE and $w_k$ satisfies \eqref{ass:w}. Then, for any initial condition $\tilde \theta_0$, the estimation error $\tilde \theta_k$ converges exponentially to a ball centered on $\theta$ with a radius proportional to the bound on $w$, i.e. for any $\tilde \theta_0$ there exist $\gamma_1,\gamma_2 >0$ such that for all $k\geq M$
\begin{align*}
|\tilde \theta_k|\leq \gamma_1 \lambda^{k/2} |\tilde \theta_0|+\gamma_2\frac{\lambda^{k/2}-1}{\lambda^{1/2}-1}\bar w.
\end{align*}\hfill$\triangle$
\end{lem}
\begin{pf}
\iftoggle{proofIFAC}
{
It is shown in \citet{JOHNSTONE1982} that the result holds for SISO systems and no disturbance. An equivalent result for the multiple output case is under review, see \citet{brggemann2020exponential}. Exponential convergence of the linear error dynamics implies BIBO stability, which gives the desired result.
}
{
The proof is divided into three parts. Part I establishes a lower bound on $P^{-1}_k$. This is used in part II to show exponential stability of the estimation error provided there does not exist any disturbance. Part III removes the latter condition and concludes with the desired statement.

\textit{Part I}

Recollect that if $B$ is symmetric, then for any matrix $A$,
\begin{align*}
ABA^{\top}\geq \lambda_{min}(B)AA^{\top}.
\end{align*}
This follows by definition of a positive definite matrix,
\begin{align*}
x^{\top}(ABA^{\top}&- \lambda_{min}(B)AA^{\top})x\\
&=x^{\top}ABA^{\top}x- \lambda_{min}(B) x^{\top}AA^{\top}x\\
&=(A^{\top}x)B(A^{\top}x)- \lambda_{min}(B) x^{\top}AA^{\top}x\\
&\geq \lambda_{min}(B) (A^{\top}x)A^{\top}x -  \lambda_{min}(B) x^{\top}AA^{\top}x\\
&=0,
\end{align*}
where we use the fact that $\lambda_{min}(B)|x|^2\leq |x|_B^2$. It follows that if $\{x_k,u_k\}$ is PE
\begin{align*}
P^{-1}_{j-1}+\dots+P^{-1}_{j+M-1}&\numeq{\ref{eq:PinvRecursion}}{\geq} \sum_{k=j}^{j+M}\varphi_{k-1}T^{-1}\varphi_{k-1}^{\top}\\
&~\geq \lambda_{min}(T^{-1})\alpha I.
\end{align*}
for all $k\geq M$. Following \citet[Lemma 1]{JOHNSTONE1982} leads to the lower bound
\begin{align}\label{eq:PinvLB}
P^{-1}_{k-1}\geq \frac{\lambda_{min}(T^{-1})\alpha(\lambda^{-1}-1)}{\lambda^{-(M+1)}-1}>0
\end{align}
for all $k\geq M$.

\textit{Part II}

By \eqref{eq:thetaHatRecursion} and assuming zero noise, one can recursively write the estimation error in \eqref{eq:thetaTilde} as
\begin{align}\label{eq:theta tilde recursion}
\tilde \theta_{k+1}=\left(I-P_{k-1}\varphi_{k}D_k^{-1}\varphi^{\top}_{k}\right)\tilde \theta_{k}.
\end{align}
Consider the Lyapunov function candidate
\begin{align}\label{eq:Wk}
W_k=\tilde \theta_k^{\top} P_{k-1}^{-1}\tilde \theta_k.
\end{align}
Using the recursions in \eqref{eq:PinvRecursion} and \eqref{eq:theta tilde recursion} then yields
\begin{align}\label{eq:Wk+1-Wk}
W_{k+1}-W_k&=\tilde \theta_{k+1}^{\top} P_{k}^{-1}\tilde \theta_{k+1}-\tilde \theta_k^{\top} P_{k-1}^{-1}\tilde \theta_k\nonumber \\
&=\tilde \theta_k^{\top}\left[(\lambda-1)P_{k-1}^{-1}-\lambda\varphi_k D_k^{-1}\varphi_k^{\top}+C\right]\tilde \theta_k,
\end{align}
where
\begin{align*}
C&=\varphi_k\left[T^{-1}-D_k^{-1}\varphi_k^{\top}P_{k-1}\varphi_kT^{-1} -\lambda D_k^{-1}\right.\nonumber\\
&\quad -T^{-1}\varphi_k^{\top} P_{k-1}\varphi_kD_k^{-1}+\lambda D_k^{-1}\varphi_k^{\top}P_{k-1}\varphi_kD_k^{-1}\nonumber \\
&\quad \left.+D^{-1}_k\varphi^{\top}_kP_{k-1}\varphi_kT^{-1}\varphi_k^{\top} P_{k-1}\varphi_kD_k^{-1}\right]\varphi_k^{\top}.
\end{align*}
We demonstrate now that $C$ is equal to the zero matrix. To this end, multiply the inner term of $C$ by $T^{1/2}$ from both sides respectively and obtain
\begin{align}\label{eq:proof T C T}
I&-T^{1/2}D_k^{-1}\varphi_k^{\top}P_{k-1}\varphi_kT^{-1/2}-\lambda T^{1/2}D_k^{-1}T^{1/2} \nonumber\\
&-T^{-1/2}\varphi_k^{\top}P_{k-1}\varphi_kD_k^{-1}T^{1/2} \nonumber\\
&+\lambda T^{1/2}D_k^{-1}\varphi_k^{\top}P_{k-1}\varphi_kD_k^{-1}T^{1/2} \nonumber\\
&+T^{1/2}D_k^{-1}\varphi_k^{\top}P_{k-1}\varphi_kT^{-1}\varphi_k^{\top}P_{k-1}\varphi_k D_k^{-1}T^{1/2}.
\end{align}
By defining
\begin{align}\label{eq:Dbar}
\bar \varphi&=\varphi_kT^{-1/2}\nonumber\\
\bar D&=T^{-1/2}D_kT^{-1/2}=\lambda I+\bar\varphi^{\top}P_{k-1}\bar \varphi
\end{align}
one has that
\begin{align*}
T^{1/2}D_k^{-1}&=(D_kT^{-1/2})^{-1}=(\lambda T^{1/2}+\varphi_k^{\top}P_{k-1}\varphi_kT^{-1/2})^{-1}\\
&=\left(T^{1/2}(\lambda+T^{-1/2}\varphi_k^{\top}P_{k-1}\varphi_kT^{-1/2})\right)^{-1}\\
&=\left(T^{1/2}(\lambda+\bar \varphi^{\top}P_{k-1}\bar \varphi)\right)^{-1}\\
&=\bar D^{-1}T^{-1/2},
\end{align*}
so that for \eqref{eq:proof T C T} it follows
\begin{align*}
I&-\bar D^{-1}\bar \varphi^{\top}P_{k-1}\bar \varphi-\lambda\bar D^{-1}\\
&-\bar \varphi^{\top}P_{k-1}\bar \varphi\bar D^{-1}+\lambda \bar D^{-1}\bar \varphi^{\top}P_{k-1}\bar \varphi\bar D^{-1}\\
&+\bar D^{-1}\bar \varphi^{\top}P_{k-1}\bar \varphi\bar \varphi^{\top}P_{k-1}\bar \varphi\bar D^{-1}.
\end{align*}
Observe that this can be reformulated as
\begin{align*}
I&-\bar D^{-1}\overbrace{\left(\bar \varphi^{\top}P_{k-1}\bar\varphi+\lambda I\right)}^{\numeq{\ref{eq:Dbar}}{=}\bar D}\\
&\quad +\bar D^{-1}\left(-\bar D\bar \varphi^{\top}P_{k-1}\bar \varphi+\lambda\bar \varphi^{\top}P_{k-1}\bar \varphi\right.\\
& \quad \quad \quad \quad \quad  \left.+\bar\varphi^{\top}P_{k-1}\bar\varphi\bar\varphi^{\top}P_{k-1}\bar\varphi\right)\bar D^{-1},
\end{align*}
which is by definition of $\bar D$ in \eqref{eq:Dbar} zero and so is $C$.

Therefore, the difference related to the Lyapunov function candidate in \eqref{eq:Wk+1-Wk}
\begin{align*}
W_{k+1}-W_k&=\tilde \theta_k^{\top}\left[(\lambda-1)P_{k-1}^{-1}-\lambda\varphi_k D_k^{-1}\varphi_k^{\top}\right]\tilde \theta_k\\
&\leq (\lambda-1)\tilde \theta_k^{\top}P_{k-1}^{-1}\tilde \theta_k\\
&=(\lambda-1)W_k
\end{align*}
so that
\begin{align*}
W_{k+1}\leq\lambda W_k\leq\lambda^{k+1}W_0=\lambda^{k+1}\tilde \theta_0^{\top}P_{-1}^{-1}\tilde \theta_0.
\end{align*}
Finally, combining this inequality with the definition of $W_k$ in \eqref{eq:Wk} and the lower bound of $P_{k-1}^{-1}$ in \eqref{eq:PinvLB} leads to
\begin{align*}
|\tilde \theta_k|^2\leq \underbrace{\frac{\lambda^{-(M+1)}-1}{\lambda_{min}(T^{-1})\alpha(\lambda^{-1}-1)}\lambda_{max}\left(P_{-1}^{-1}\right)}_{\eqqcolon \gamma_F}\lambda^k|\tilde \theta_k|^2
\end{align*}
for all $k\geq M$, which proves exponential stability in the absence of noise.

\textit{Part III}

Consider \eqref{eq:thetaTilde} and substitute \eqref{eq:xk+1_varphi} into \eqref{eq:thetaHatRecursion}. Consequently,
\begin{align*}
\tilde \theta_k&=\overbrace{\left(I-P_{k-2}\varphi_{k-1}D_{k-1}^{-1}\varphi_{k-1}^{\top}\right)}^{\eqqcolon F_{k-1}}\tilde \theta_{k-1}\nonumber \\[1em]
&\quad \quad \quad \quad \quad \quad  -\underbrace{P_{k-2}\varphi_{k-1}D_{k-1}^{-1}}_{\eqqcolon B_{k-1}}w_{k-1}.
\end{align*}
Hence,
\begin{align}\label{eq:theta tilde proof}
\tilde \theta_k=\underbrace{\prod_{i=0}^{k-1}F_i}_{\eqqcolon\tilde F_k}\tilde \theta_0-\sum_{j=0}^{k-1}\left(\prod_{i=j+1}^{k-1}F_i\right)B_jw_j.
\end{align}
From part II one deduces that for all $\tilde \theta_0$ and $k\geq M$,
\begin{align*}
|\tilde F_k\tilde \theta_0|^2\leq \gamma_F\lambda^k|\tilde\theta_0|^2.
\end{align*}
Taking the supremum leads to
\begin{align*}
\tilde \lambda_{k, max}\coloneqq\lambda_{max}\left(\tilde F_k^{\top}\tilde F_k\right)\leq \gamma_F \lambda^k.
\end{align*}

Hence, with
\begin{align*}
|\tilde F_k\tilde \theta_0|^2\leq |\tilde F_k|^2 \,|\tilde \theta_0|^2= \tilde\lambda_{k,max} \,|\tilde \theta_0|^2\leq \gamma_F\lambda^k|\tilde \theta_0|^2,
\end{align*}
and letting $\gamma_1\coloneqq\sqrt{\gamma_F}$, the error dynamics in \eqref{eq:theta tilde proof} are bounded as follows.
\begin{align*}
|\tilde \theta_k|&\leq \gamma_1 \lambda^{k/2} |\tilde \theta_0| +\sum_{j=0}^{k-1}\left\rvert\left(\prod_{i=j+1}^{k-1}F_i\right)\right\rvert\,\left\rvert B_jw_j\right\rvert\nonumber\\
&\leq \gamma_1 \lambda^{k/2} |\tilde \theta_0|+\sum_{j=0}^{k-1}\gamma_1\lambda^{(k-j-1)/2}\,\left\rvert B_jw_j\right\rvert\nonumber \\
& \leq \gamma_1 \lambda^{k/2} |\tilde \theta_0|+\underbrace{\gamma_1\gamma_B}_{\eqqcolon\gamma_2}\frac{\lambda^{k/2}-1}{\lambda^{1/2}-1}\bar w,
\end{align*}
where
\begin{align*}
\gamma_B\coloneqq \frac{\lambda^{-(S+1)}-1}{\lambda_{min}(T^{-1})\alpha(\lambda^{-1}-1)}\beta^{1/2}\lambda^{-1}\lambda_{max}(T^{-1}).
\end{align*}
The constant $\gamma_B$ is a consequence of the upper bound on $P_k$ from part II and the definition of persistent excitation.
}
\hfill$\qed$
\end{pf}

We have thus shown that under the assumption of a bounded disturbance and a PE sequence, the estimate converges exponentially to the actual parameter without noise, or in the case of a bounded disturbance, to a neighborhood whose size depends on the bound of the disturbance. Exponential convergence is decisive for the preservation of a PE closed loop sequence, as disclosed in the next section, where we combine previous results.
\section{Periodic adaptive model predictive control}
All the local results above share a common concept. That is, in a utopian world with suitable initial conditions, perfect knowledge of the uncertainty and under sufficient conditions, a PE closed loop is guaranteed. Gradually watering down these conditions by contemplating sufficiently small neighborhoods has been shown not to affect the substance of the initial statement about the PE closed loop sequence, provided we carry along suitable smoothness and regularity assumptions. Consistent with this strategy, this section focuses on the the estimation error and its interplay with the neighborhoods introduced before. In this fashion, by noting that if the bound on the estimation error implies a neighborhood for which a PE closed loop sequence exists, then we achieve a PE closed loop sequence despite uncertainty.
\subsection{Convergence under bounded noise}
The following theorem states that under sufficient conditions, if the disturbance is bounded and the initial state and the initial parameter estimate are within a neighborhood of the periodic PE reference trajectory and the true parameter, respectively, then the estimation error and the closed loop tracking error exponentially converge to a neighborhood around the reference trajectory and the true parameter, respectively.
\begin{thm}\label{thm:x_theta_prac_conv}
Suppose Assumption \ref{ass:fC2}, \ref{ass:posDefHessian} and \ref{ass:localIncrementalStabilizability} hold. Let the control input be derived by the optimization problem in \eqref{eq:Vn} with $\theta$ substituted by $\hat\theta_0$ for $k<M$ and $\hat \theta_k$ given by the recursion in \eqref{eq:thetaHatRecursion} for $k\geq M$. Then, for any feasible PE reference trajectory there exist $c_{w,k}, c_{\theta}, \bar w$ and a sufficiently large horizon $N$ such that for all $|x_0-x_r(0)|\leq c_{w,k}$, $|w_k|\leq \bar w, |\tilde \theta_0|\leq c_{\theta},$
\begin{align*}
x_k-x_r(k)&\to \Z_{RPI}\\
|\tilde \theta_k|&\to\frac{\gamma_2}{1-\lambda^{1/2}}\bar w.
\end{align*}as $k\to\infty$.\hfill $\triangle$
\end{thm}
\begin{pf}
Let $c_{x,1}, \bar w_1, c_{\theta,1}>0$ such that Corollary \ref{thm:PEAfterKEpsilonTheta} holds. Then, with Lemma \ref{lem:tilde theta convergence}, let $\bar w_2, c_{\theta,2}>0$ such that $|\tilde \theta_0|\leq c_{\theta,2}$ and $|w_k|<\bar w_2$ imply $|\tilde \theta_k|\leq c_{\theta,1}$ for all $k\geq M$. By selecting $\bar w=\min\{\bar w_1,\bar w_2\}$ and $c_{\theta}=\min\{c_{\theta,1}, c_{\theta,2}\}$, the conclusion follows from Corollary \ref{thm:PEAfterKEpsilonTheta} and Lemma \ref{lem:tilde theta convergence}.\hfill$\qed$
\end{pf}
Observe that the aforementioned theorem relies on a periodic PE reference trajectory, also depending on the uncertain parameter. By continuity arguments, an equivalent statement holds for a periodic PE reference generated with an initial parameter estimate in a sufficiently small neighborhood of the true parameter. However, persistence of excitation and feasibility of the reference trajectory is generally not ensured for all initial estimates which may deviate substantially from the true parameter. The same obstacle may occur if the reference trajectory is updated online using the current estimate. Furthermore, note that the convergence result is only local with respect to the uncertain parameter and the initial state. The upshot is that the exponential convergence result does not inherit the usual drawback of unknown transient performance common in adaptive control, see e.g. \citet{ariyur2014}. The convergence result for the uncertain and perturbed system is demonstrated numerically in the next section for a non-infinitesimal neighborhood of initial conditions about their nominal values.
\section{Simulation example}
Consider the nonlinear scalar system from \citet{HOVD2004119},
\begin{align}\label{eq:ExSys}
x_{k+1}&=\theta_1x_k+\theta_2 x_ku_k+u_k+w_k,
\end{align}
or equivalently, 
\begin{align*}
x_{k+1}=f_0(x_k,u_k)+\varphi^{\top}\theta +w_k,
\end{align*}
where
\begin{align}\label{eq:f0ThetaPhi}
f_0(x_k,u_k)&=u_k,\nonumber \\
\theta&=\begin{bmatrix}\theta_1 & \theta_2\end{bmatrix}^{\top},\\
\varphi_k&=\begin{bmatrix}f_1(x_k) & f_2(x_k,u_k)\end{bmatrix}^{\top},\nonumber
\end{align}
with
\begin{align}\label{eq:f1f2}
f_1(x_k)=x_k, \quad f_2(x_k,u_k)=x_ku_k.
\end{align}
The parameters $\theta_1=1.1$ and $\theta_2=0.1$ are unknown and the noise $|w_k|<0.2$. The main objective is to regulate the state to a steady tuple $(x_s,u_s)$, which presumes an accurate estimate of the unknown parameter $\theta$. Therefore, we first generate a PE feasible periodic reference trajectory around the steady tuple using the results from Section \ref{sec:PErefTraj}. Then, we employ the MPC from \eqref{eq:Vn} with an estimate given by the recursive least squares in \eqref{eq:thetaHatRecursion} - \eqref{eq:Pk_recursion} with a forgetting factor $\lambda=0.9$ and weight $T=1$. 

\subsection{Regulate to steady tuple outside the origin}
Consider the steady tuple $(x_s,u_s)=(1,-0.09)$ around which we want to generate a PE reference trajectory. Observe that $f_j,\,j\in\{0,1,2\},$ in \eqref{eq:f0ThetaPhi} and \eqref{eq:f1f2} are twice continuously differentiable and that by \eqref{eq:AB} and \eqref{eq:CsDs},
\begin{align*}
A&=1.09, &\quad B&=0.99,\\
A_1&=1, &\quad B_1&=0,\\
A_2&=-0.09, &\quad B_2&=1,\\
C_1&=\begin{bmatrix}
1 & -.09
\end{bmatrix}^{\top}, &\quad D_1&=\begin{bmatrix}
0& 1
\end{bmatrix}^{\top}.
\end{align*}
Hence, as $(A,B)=(1.09,0.99)$ is controllable and $\lambda(A)=1.09$, by Lemma \ref{lem:perReachRef}, for any positive $M$ there exists a feasible period-$M$ reference trajectory. Further, it is easy to see that the McMillan degree $d_1=1$, and thus, the output reachability matrix
\begin{align*}
\begin{bmatrix}
D_1 & C_1B
\end{bmatrix}=\begin{bmatrix}
0 & 0.99\\
1 &-0.09 
\end{bmatrix}
\end{align*}
which has full rank. As a result, by Proposition \ref{prop:uPEimpliesPhiPE}, it is sufficient to generate an input sequence which satisfies the persistence of excitation condition therein. Consider
\begin{align*}
\delta u_r(k)=\bar u\sin\left(\frac{2\pi}{M}k\right),
\end{align*}
where $\bar u=0.3$, and let $M=4$. Then, the condition holds since for all $j$,
\begin{align*}
\alpha_u\leq\sum_{i=j}^{j+2}\delta u_r(i)^2\leq \beta_u,
\end{align*}
where
$\alpha_u=(\bar u\sin(2\pi/M))^2$ and $\beta_u=(\bar u(M-d_1))^2$. Thus, the corresponding feasible period-$M$ sequence $\{x_r(k),\allowbreak u_r(k)\}$ is PE. By solving \eqref{eq:x_r0} numerically, we compute the corresponding $x_r(0)=0.91$.

\begin{figure}[ht]
    \centering
    \includegraphics[width=\columnwidth]{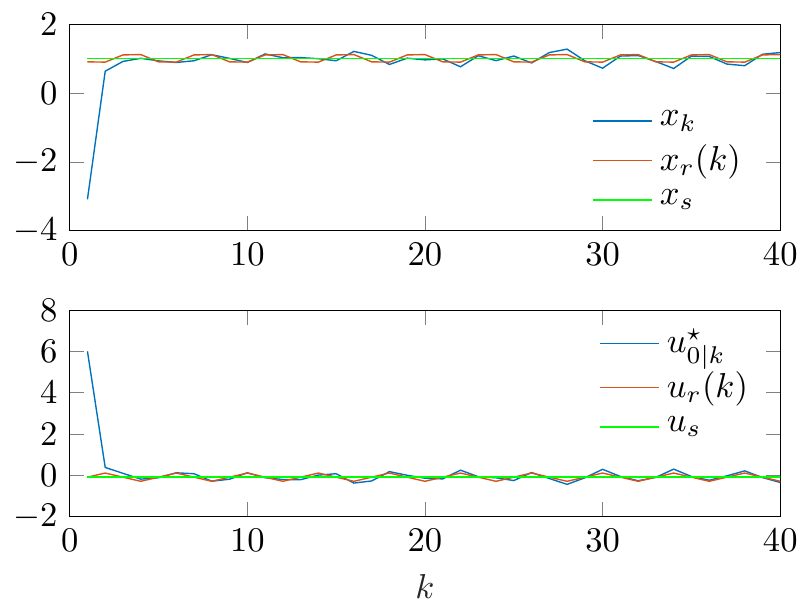} 
 \caption{Closed loop with noised $|w_k|\leq 0.2$ versus PE reference trajectory}
    \label{fig:xu}
\end{figure}

\begin{figure}[ht]
    \centering
     \includegraphics[width=\columnwidth]{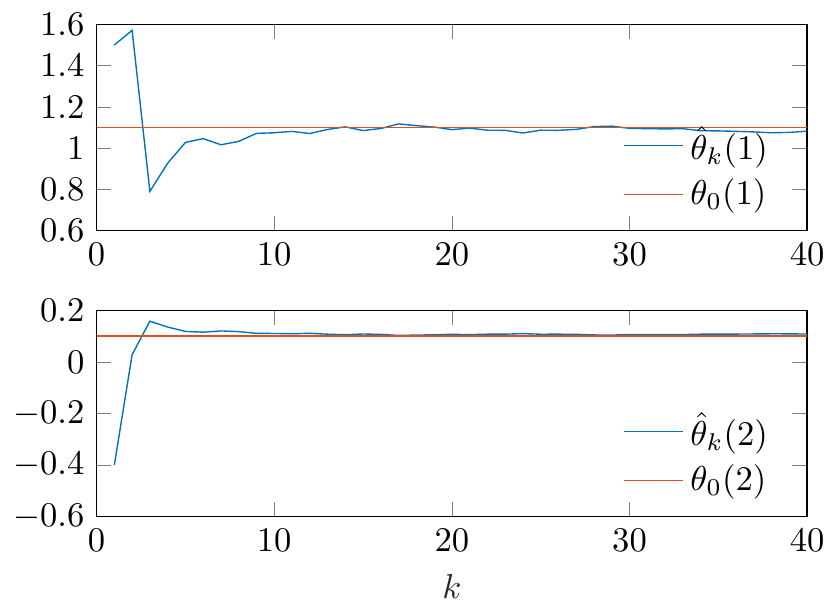} 
 \caption{Parameter estimate given measurement noise $|w_k|\leq0.2$.}
    \label{fig:theta hat}
\end{figure}

\begin{figure}[ht]
    \centering
  \includegraphics[width=\columnwidth]{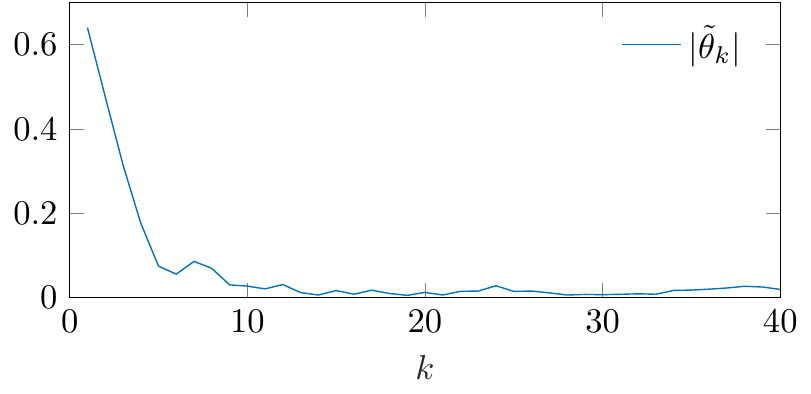} 
 \caption{Euclidean norm of the estimation error given measurement noise $|w_k|\leq0.2$.}
    \label{fig:theta tilde}
\end{figure}

Having generated a PE period-$M$ sequence, we next elaborate on how the assumptions of Theorem \ref{thm:x_theta_prac_conv} are satisfied. Twice continuous differentiability trivially holds for all $f_i,\,i\in\{1,2,3\}$. Assumption \ref{ass:localIncrementalStabilizability} holds by letting
\begin{align*}
\kappa(x_k,x_r(k),u_r(k))&=\frac{1}{\theta_2x_k+1}K(x_k-x_r(k))\\
& \quad \quad \quad \quad \quad \quad \quad +u_r(k)(\theta_2x_r(k)+1)\\
V_{\delta}&=|x_k-x_r(k)|^{2}_{P},
\end{align*}
where $P$ and $K$ relate to the discrete-time infinite-horizon linear quadratic regulator using common notation. Lastly, Assumption \ref{ass:posDefHessian} is numerically verified. We subsequently present the simulation of the closed loop driven by the MPC in \eqref{eq:Vn} with weights $Q=6, R=0.1$ and horizon $N=4$. The disturbance has a uniform distribution in the interval $[-0.2,0.2]$. The figures below are based on an initial estimate $\hat \theta_0=[1.5~-0.4]^{\top}$. Figure \ref{fig:xu} depicts a fast convergence of the closed loop to a small neighborhood of the reference trajectory. The error between the reference trajectory and the closed loop can be explained by the initial parameter uncertainty and the noise. Still, the system can be regulated to a neighborhood of the steady tuple $(1,-0.09)$ despite noise and parameter uncertainties.
The uncertain parameter estimate is visualized in Figure \ref{fig:theta hat}. Akin to the closed-loop trajectory, the estimate converges to a small ball centered on the true parameter, after which the estimates continuously move around within this ball. A similar pattern can be observed in Figure \ref{fig:theta tilde} in which the norm of the estimation error is plotted.
The estimation error converges exponentially to a neighborhood of the origin and remains there.
As the analysis in the previous sections is of local and depends on the selected steady tuple, the assumptions are generally \emph{not necessary} but \emph{sufficient}.
\subsection{Sufficiency of imposed assumptions}
There are various cases when the assumptions are not satisfied, but the reference trajectory is still PE and the closed loop still converges. For example, as in \citet{SvenBobIFAC2020}, suppose we want to regulate the system in \eqref{eq:ExSys} to the origin, i.e. 
\begin{align*}
(x_s,u_s)=(0,0),
\end{align*}
and additionally suppose that the true uncertain parameter 
\begin{align*}
\theta^{\top}=\begin{bmatrix}
\theta_1 & \theta_2
\end{bmatrix}=\begin{bmatrix}
1 & 0.1
\end{bmatrix}.
\end{align*}
The related matrices introduced earlier are
\begin{align*}
A&=1, &\quad B&=1,\\
A_1&=1, &\quad B_1&=0,\\
A_2&=0, &\quad B_2&=0,\\
C_1&=\begin{bmatrix}
1 & 0
\end{bmatrix}^{\top}, &\quad D_1&=\begin{bmatrix}
0& 0
\end{bmatrix}^{\top}.
\end{align*}
Then, the assumptions of Lemma \ref{lem:perReachRef} does not hold as the eigenvalue 
\begin{align*}
\lambda(A)=1.
\end{align*}
Moreover, nor the condition of output reachability holds as with a McMillan degree of $d_1=1$,
\begin{align*}
\begin{bmatrix}
D_1 & C_1B
\end{bmatrix}=\begin{bmatrix}
0 & 1\\
0 & 0
\end{bmatrix}
\end{align*}
has not full rank. 

Alternatively, we may generate the PE period-$M$ reference trajectory adhoc by solving an optimization problem with a PE constraint:
\begin{align}\label{eq:optProbRefTraj}
(\bar x_{r},\bar u_{r})& =\argmin_{\substack{\{x_0,x_{1}\}\\\{u_0,u_{1}\}}} \frac{1}{N} \sum_{i=0}^{N-1}|x_{i|k}|_Q^2+|u_{i|k}|_R^2,\nonumber \\
s.t.  ~x_{k+1}&=f(x_{k},u_{k})\nonumber \\
x_{N}&=x_0\\
\alpha I &\leq \sum_{i=0}^{N-1}\varphi_i^{\top}\varphi_i\leq \beta I,\nonumber
\end{align}
where $\alpha=0.1, \beta=0.3,Q=6, R=0.1, N=2$. The associated reference trajectory can be defined by
\begin{align*}
x_{r}(k)&=\bar x_{r}(k \mod 2)\\
u_{r}(k)&=\bar u_{r}(k \mod 2 ).
\end{align*}
We chose $N=2$ since it results in a low cost relative to other small integer values. The solution of the minimization problem guarantees a PE period-$2$ reference trajectory. It also ensures that the reference values are in the neighborhood of the steady state. The optimization problem is solved within seconds on a regular laptop, by the interior point algorithm in MATLAB.
Under otherwise equal conditions as in the previous example, the subsequent figures illustrate the closed loop and the related estimates under the influence of the MPC in \eqref{eq:Vn} and the estimator governed by \eqref{eq:thetaHatRecursion} - \eqref{eq:Pk_recursion}.
\begin{figure}[ht]
    \centering
\includegraphics[width=\columnwidth]{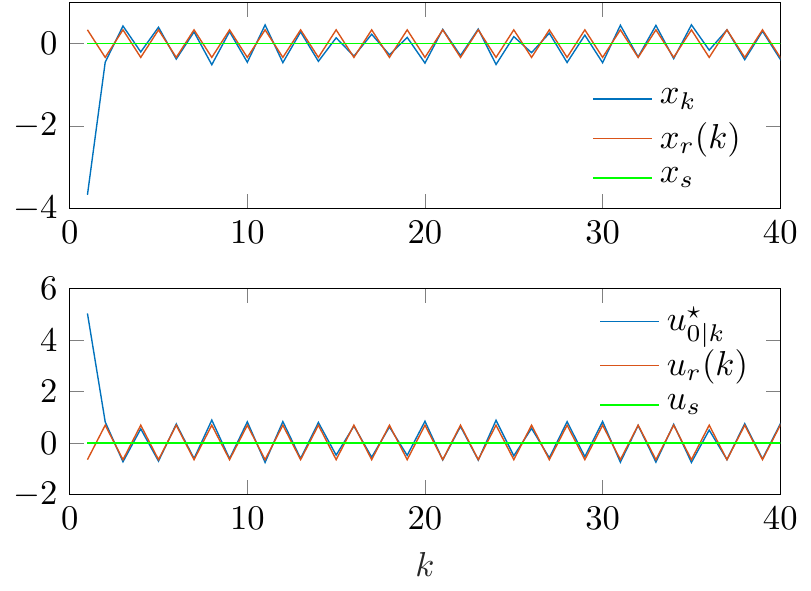} 
 \caption{Closed loop with reference trajectory generated by optimization problem in \eqref{eq:optProbRefTraj}.}
    \label{fig:xuOpt}
\end{figure}
Figure \ref{fig:xuOpt} shows that the closed loop converges to a small neighborhood around the given reference trajectory. The size of the neighborhood correlates with the bound on the noise. Also notice that the optimization problem in \eqref{eq:optProbRefTraj} results in a more aggressive reference trajectory with higher amplitudes, which may also be caused by the shorter period and by the different state tuple targeted for regulation.
\begin{figure}[ht]
    \centering
 \includegraphics[width=\columnwidth]{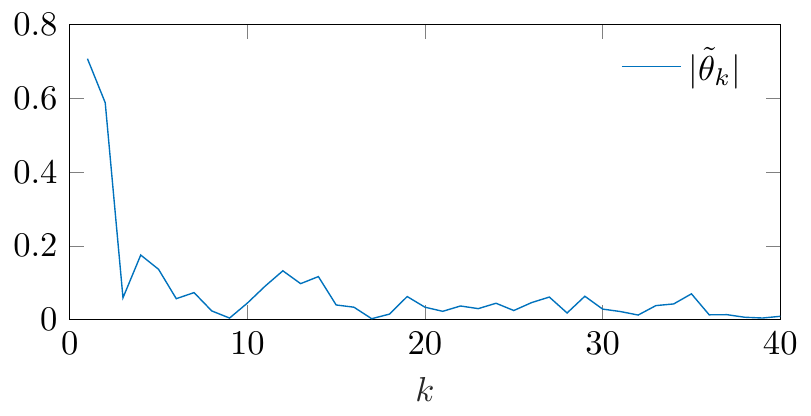} 
 \caption{Euclidean norm of the estimation error corresponding to reference trajectory generated by optimization problem in \eqref{eq:optProbRefTraj}.}
    \label{fig:theta tildeOpt}
\end{figure}
Similarly to the first example, the estimation error also converges to a neighborhood of zero, whereby the fluctuating distance to the origin may be a result from the measurement noise. As a concluding remark, this example shows that it  may still be possible to create a PE feasible periodic reference trajectory, although the assumptions on the system linearized around the steady tuple may prohibit the application of some of our theory. It also stresses that the steady tuple must be chosen carefully in order to meet all sufficient conditions presented in this work.

\section{Conclusion}
This work presents a constructive proof for the existence of a PE reference trajectory and hence a simple procedure for the creation of such a trajectory. This enables PE closed loop sequence by only looking forward in time despite disturbances, uncertainties and the MPC's nature of a receding horizon implementation. Additionally, due to computations offline, the online optimization problem does not complicate. The theory is supported by two simulation examples which underpin the sufficient nature of our assumptions. Future work will revolve around the recomputation of the reference trajectory given more accurate estimates over time, and its implications for stability, feasibility and persistence of excitation.

\bibliographystyle{humannat}
\bibliography{ifacconf}             
                                                   






%
%
\end{document}